\documentclass[10pt,conference]{IEEEtran}
\IEEEoverridecommandlockouts
\usepackage{lineno}
\usepackage{hyperref}
\usepackage{cite}
\usepackage{amsmath,amssymb,amsfonts,cases} 
\usepackage{amsmath}
\usepackage{amsthm}
\DeclareMathOperator*{\argmax}{argmax}

\usepackage{graphicx}
\usepackage{textcomp}
\usepackage{xcolor}
\usepackage{graphicx}
\usepackage{float}
\usepackage{subfigure}
\usepackage{amsmath,mleftright}
\usepackage{amsfonts,amssymb}
\usepackage{mathrsfs}
\usepackage{mathtools}
\usepackage{algorithm}
\usepackage{algorithmic}
\usepackage{bm}
\usepackage{multirow}
\usepackage{array}
\usepackage{amssymb}
\usepackage{amsmath}
\usepackage{cite}
\usepackage{url}
\usepackage{xcolor}
\usepackage{cite,graphicx,amsmath,amssymb}
\usepackage{subfigure}
\usepackage{fancyhdr}
\usepackage{mdwmath}
\usepackage{mdwtab}
\usepackage{caption}
\usepackage{amsthm}
\usepackage{setspace}
\usepackage{bm}
\usepackage{mathtools}
\usepackage{dsfont}
\usepackage{bbm}
\usepackage{framed}
\newtheorem{remark}{Remark}
\newtheorem{theorem}{Theorem}

\newtheorem{lemma}{Lemma}

\newtheorem{corollary}{Corollary}

\makeatletter
\newcommand{\biggg}{\bBigg@{3}}
\newcommand{\Biggg}{\bBigg@{3.5}}
\makeatother
\makeatletter
\allowdisplaybreaks[4]
\renewcommand{\maketag@@@}[1]{\hbox{\m@th\normalsize\normalfont#1}}%
\makeatother
\def\BibTeX{{\rm B\kern-.05em{\sc i\kern-.025em b}\kern-.08em
    T\kern-.1667em\lower.7ex\hbox{E}\kern-.125emX}}
    \expandafter\def\expandafter\normalsize\expandafter{%
    \normalsize%
    \setlength\abovedisplayskip{4pt}%
    \setlength\belowdisplayskip{4pt}%
    \setlength\abovedisplayshortskip{2pt}%
    \setlength\belowdisplayshortskip{2pt}%
}
\begin{document}
\title{On the Performance of Tri-Hybrid Beamforming Using Pinching Antennas}
\author{\IEEEauthorblockN{Zhenqiao Cheng$^{\dag}$, Chongjun~Ouyang$^{\star}$, and Nicola~Marchetti$^{\ddag}$}
$^\dag$6G Research Centre, China Telecom Beijing Research Institute, Beijing, 102209, China\\
$^{\star}$Queen Mary University of London, London, U.K. $^{\ddag}$Trinity College Dublin, Dublin, Ireland}
\maketitle
\begin{abstract}
The Pinching-Antenna System (PASS) reconfigures wireless channels through \emph{pinching beamforming}, in which the active positions of pinching antennas (PAs) along dielectric waveguides are optimized to shape the radiation pattern. This article investigates the performance of PASS-enabled tri-hybrid beamforming, where pinched waveguides are integrated with a hybrid digital-analog beamformer to mitigate path loss and enhance spectral efficiency. The channel capacity of the proposed system is characterized by deriving the optimal tri-hybrid beamformer at both the digital and analog domains, as well as the optimal placement of PAs. Closed-form upper and lower bounds of the channel capacity are obtained, leading to a capacity scaling law with respect to the number of PAs. Numerical results verify the tightness of the derived bounds and demonstrate that applying PASS to tri-hybrid beamforming yields a significant performance gain over conventional hybrid beamforming under the same number of radio-frequency chains.
\end{abstract}
\section{Introduction}
Over the past three decades, multiple-input multiple-output (MIMO) technology has fundamentally transformed wireless communications. Recently, MIMO research has entered a new phase that focuses on \emph{reconfigurable antennas} \cite{you2025next}, which complement conventional digital and analog beamforming as an additional adaptive layer. By adjusting their radiation patterns in real time, these antennas enable flexible electromagnetic (EM) beam control and channel reconfiguration. Reconfigurable antennas are now regarded as a key component of the emerging \emph{tri-hybrid beamforming architecture}, which is expected to play a pivotal role in shaping sixth-generation (6G) wireless networks.

Representative examples of reconfigurable antennas for supporting tri-hybrid beamforming include dynamic metasurface antennas, fluid antennas, and movable antennas \cite{heath2025tri,cheng2024enabling,cheng2024sum}. Although these approaches provide notable performance gains, their reconfiguration capability is typically confined to apertures spanning only a few to several tens of wavelengths. Such a scale remains insufficient to overcome two fundamental constraints of wave propagation, particularly in the high-frequency bands envisioned for 6G, such as the 7-24 GHz upper mid-bands: \emph{severe large-scale free-space path loss} and \emph{high susceptibility to blockages} \cite{liu2025pinching}.

To address these challenges, NTT DOCOMO introduced the pinching-antenna system (PASS) in 2021 together with a working prototype \cite{suzuki2022pinching}. PASS employs low-loss dielectric waveguides as signal conduits, to which small dielectric particles, termed pinching antennas (PAs), are attached. Each PA radiates signals into free space or collects incident signals into the waveguide. By adjusting the PA positions, one can control the phase and amplitude of the radiated signals, thereby forming flexible beampatterns and enabling \emph{pinching beamforming} \cite{liu2025pinching}. Unlike conventional reconfigurable antennas, PASS supports arbitrarily long waveguides, which allows the deployment of PAs in close proximity to users. This adaptability establishes strong line-of-sight (LoS) links and mitigates both large-scale path loss and blockage effects  \cite{liu2025pinching}. In summary, PASS combines low propagation loss with high spatial flexibility in signal control.

Early field tests conducted by NTT DOCOMO confirmed the feasibility of using pinched waveguides to enhance both network coverage and throughput \cite{suzuki2022pinching}. These results have stimulated growing research interest in exploiting PASS for communication performance improvement. Several theoretical studies have demonstrated the superiority of PASS over conventional fixed-antenna systems and existing reconfigurable-antenna architectures \cite{ding2024flexible,ouyang2025array}. In addition, various pinching beamforming methods have been proposed to optimize the placement of PAs \cite{xu2024rate,ouyang2025capacity,wang2025modeling,bereyhi2025mimo}. However, the integration of PASS into the tri-hybrid beamforming framework, an emerging 6G architecture that employs reconfigurable antennas, such as pinched waveguides, as an outer layer complementing conventional hybrid digital-analog beamforming \cite{heath2025tri}, has not yet been explored. Such an integration can improve spectral efficiency while maintaining low cost and energy consumption \cite{castellanos2025embracing}.

\begin{figure}[!t]
\centering
\includegraphics[height=0.08\textwidth]{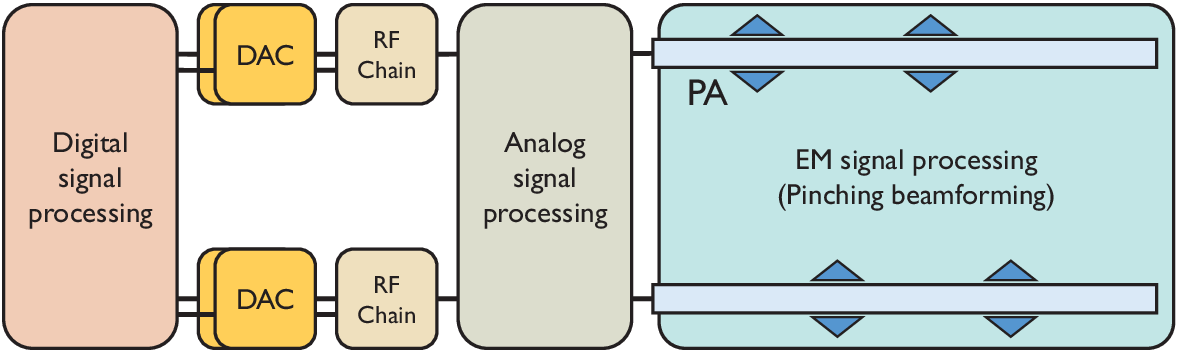}
\caption{Illustration of the tri-hybrid beamforming architecture.}
\label{System_Model}
\vspace{-15pt}
\end{figure}

To address this research gap, this work takes the first step toward analyzing the performance of tri-hybrid beamforming with PAs. We propose a novel tri-hybrid beamforming architecture, where the signal precoded by conventional hybrid digital-analog beamforming is further processed through EM pinching beamforming along pinched waveguides to mitigate path loss; see {\figurename} {\ref{System_Model}}. We characterize the channel capacity of the proposed architecture through an accurate analytical approximation and tight upper and lower bounds, from which the capacity scaling law with respect to the number of PAs is derived. Numerical results validate the analysis and show that the proposed tri-hybrid beamforming architecture achieves a substantial performance gain over traditional hybrid beamforming, establishing it as a promising MIMO architecture for future 6G networks.
\section{System Model and Problem Formulation}
\subsection{Signal Model}
\begin{figure}[!t]
\centering
\includegraphics[height=0.08\textwidth]{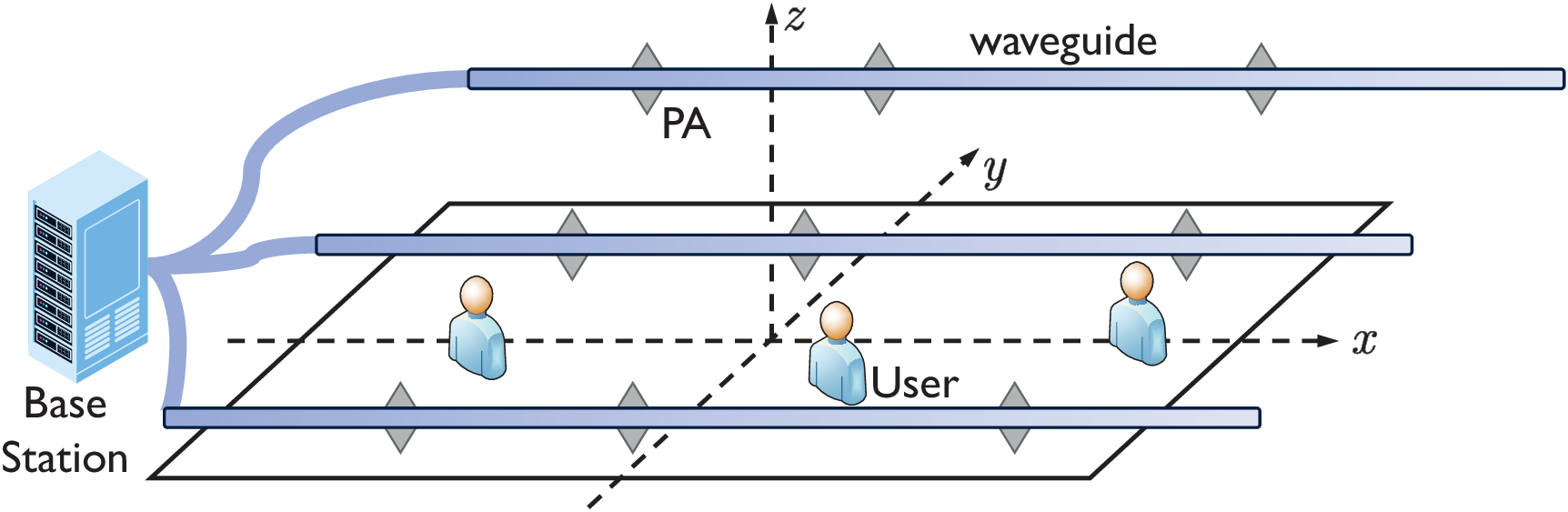}
\caption{Illustration of the system model.}
\label{System_Model1}
\vspace{-15pt}
\end{figure}

In a narrowband downlink single-cell multiuser system, a tri-hybrid base station (BS) equipped with $N_{\rm{rf}}$ transmit radio-frequency (RF) chains serves $K$ single-antenna users, as depicted in {\figurename} {\ref{System_Model1}}. The system operates under time-division multiple access (TDMA), where each user is allocated a distinct time slot and receives a single data stream during its transmission period. For brevity, we focus on a single typical user and omit the user index in the following derivations.

The tri-hybrid beamforming architecture operates in three stages. First, the BS applies an $N_{\rm{rf}}\times 1$ digital precoder ${\mathbf{w}}_{\rm{dig}}\in{\mathbb{C}}^{N_{\rm{rf}}\times1}$ at baseband to process the data streams. The signal then passes through $N_{\rm{rf}}$ RF chains, which up-convert it to the carrier frequency $f_{\rm{c}}$. Next, the BS applies an $M\times N_{\rm{rf}}$ analog precoder ${\mathbf{W}}_{\rm{ana}}$, which is implemented with analog phase shifters. The architecture imposes a unit-modulus constraint on each entry of ${\mathbf{W}}_{\rm{ana}}$. Finally, the resulting signal propagates through $M$ pinched waveguides, each allowing EM waves to radiate into free space from the $N$ activated PAs. 

The received signal at the user can be written as follows:
\begin{align}
y={\mathbf{h}}^{T}{\mathbf{G}}{\mathbf{W}}_{\rm{ana}}{\mathbf{w}}_{\rm{dig}}{{s}}+n,
\end{align}
where $s\sim{\mathcal{CN}}(0,1)$ denotes the normalized data symbol intended for the user, and $n\sim{\mathcal{CN}}(0,\sigma^2)$ represents the additive white Gaussian noise (AWGN) with variance $\sigma^2$. The vector ${\mathbf{h}}\in{\mathbb{C}}^{MN\times1}$ denotes the channel from all pinched waveguides to the user, while the block-diagonal matrix ${\mathbf{G}}\triangleq{\rm{blkdiag}}({\mathbf{g}}_{1},\ldots,{\mathbf{g}}_{M})\in{\mathbb{C}}^{MN\times M}$ formed by $\{{\mathbf{g}}_{m}\}_{m=1}^{M}$ captures the in-waveguide signal propagation effects, where ${\mathbf{g}}_{m}\in{\mathbb{C}}^{N\times1}$ models the propagation within the $m$th waveguide. Assuming that all waveguides are aligned parallel to the $x$-axis at a uniform height $H$, the entries of ${\mathbf{g}}_m$ are given by $[{\mathbf{g}}_{m}]_n=\sqrt{{\alpha_{m,n}}}{\rm{e}}^{-{\rm{j}}\frac{2\pi}{\lambda_{\rm{g}}}\lVert{\bm\psi}_{m,n}-{\bm\psi}_{m,0}\rVert}$ for $m\in{\mathcal{M}}\triangleq\{1,\ldots,M\}$ and $n\in{\mathcal{N}}\triangleq\{1,\ldots,N\}$, where the parameters are defined as follows:
\begin{itemize}
  \item  ${\bm\psi}_{m,0}=[x_{0},y_{m},H]^T\in{\mathbb{R}}^{3\times1}$: feed-point location of the $m$th waveguide.
  \item  ${\bm\psi}_{m,n}=[x_{m,n},y_{m},H]^T\in{\mathbb{R}}^{3\times1}$: position of the $n$th activated PA in the $m$th waveguide, i.e., the $(m,n)$th antenna, satisfying $x_{m,n}\geq x_{0}$.
  \item  $\lambda_{\rm{g}}=\frac{\lambda}{n_{\rm{eff}}}$: guided wavelength, where $\lambda$ denotes the free-space wavelength and $n_{\rm{eff}}$ is the effective refractive index of the dielectric waveguide.
  \item $\frac{2\pi}{\lambda_{\rm{g}}}\lVert{\bm\psi}_{m,n}-{\bm\psi}_{m,0}\rVert=\frac{2\pi}{\lambda_{\rm{g}}}(x_{m,n}-x_{0})$: in-waveguide phase shift for the $(m,n)$th PA.
  \item $\alpha_{m,n}=\frac{1}{N}10^{-\frac{\kappa\lVert{\bm\psi}_{m,n}-{\bm\psi}_{m,0}\rVert}{10}}=\frac{1}{N}10^{-\frac{\kappa(x_{m,n}-x_{0})}{10}}$: average in-waveguide attenuation factor for the $(m,n)$th PA, where $\kappa$ (in dB/m) represents the average propagation loss along the dielectric waveguide \cite{wang2024antenna}. 
\end{itemize}
\subsection{Channel Model}
Given the promising potential of PASS in high-frequency bands \cite{suzuki2022pinching}, where LoS propagation typically dominates, we adopt a free-space LoS channel model to analyze the theoretical performance limits of the proposed tri-hybrid beamforming architecture. Based on the spherical-wave channel model, the LoS channel coefficient between the $(m,n)$th PA and the user is expressed as follows: \cite{ouyang2024primer,liu2023near}
\begin{align}
h_{m,n}=\frac{\sqrt{\eta}{\rm{e}}^{-{\rm{j}}\frac{2\pi}{\lambda}\lVert{\bm\psi}_{m,n}-{\mathbf{r}}\rVert}}{\lVert{\bm\psi}_{m,n}-{\mathbf{r}}\rVert}
=\frac{\sqrt{\eta}{\rm{e}}^{-{\rm{j}}\frac{2\pi}{\lambda}r_{m,n}}}{r_{m,n}},
\end{align}
where $\eta=\frac{c^2}{16\pi^2f_{\rm{c}}^2}$, $c$ is the speed of light, $f_{\rm{c}}$ is the carrier frequency, and $\lambda$ is the corresponding wavelength. The user's position is denoted by ${\mathbf{r}}=[x_{{\rm{u}}},y_{{\rm{u}}},0]^T\in{\mathbb{R}}^{3\times1}$, and $r_{m,n}\triangleq\lVert{\bm\psi}_{m,n}-{\mathbf{r}}\rVert=\sqrt{(x_{m,n}-x_{{\rm{u}}})^2+H_{m}^2}$, where $H_{m}\triangleq\sqrt{(y_m-y_{{\rm{u}}})^2+H^2}$ represents the effective elevation distance between the user and the $m$th waveguide. Consequently, the overall channel vector from all pinched waveguides to the user can be expressed as ${\mathbf{h}}^T=[{\mathbf{h}}_{1}^T,\ldots,{\mathbf{h}}_{M}^T]$, where ${\mathbf{h}}_{m}\triangleq[h_{m,1},\ldots,h_{m,N}]^T$ denotes the channel vector corresponding to the $m$th waveguide.
\subsection{Problem Formulation}
The received signal-to-noise ratio (SNR) at the user for decoding the desired data symbol is expressed as follows:
\begin{align}\label{SINR_Basic}
{\rm{SNR}}=\lvert{\mathbf{h}}^{T}{\mathbf{G}}{\mathbf{W}}_{\rm{ana}}{\mathbf{w}}_{{\rm{dig}}}\rvert^2/{\sigma^2}.
\end{align}
It is noted that both ${\mathbf{h}}$ and ${\mathbf{G}}$ are functions of the PA positions, which are parameterized by the pinching beamformer ${\mathbf{W}}_{\rm{pin}}\in{\mathbb{R}}^{M\times N}$, where $[{\mathbf{W}}_{\rm{pin}}]_{m,n}=x_{m,n}$ for $m\in{\mathcal{M}}$ and $n\in{\mathcal{N}}$. By optimally determining the PA positions, the transceiver distance to the user can be minimized, thereby substantially reducing the path loss. This spatial adaptability enables the tri-hybrid beamforming architecture to achieve superior performance compared with conventional hybrid schemes. In this framework, pinching beamforming serves as an \emph{outer-layer control mechanism} that complements both digital and analog beamforming, offering additional spatial degrees of freedom.

The objective of this work is to characterize the channel capacity by jointly optimizing digital, analog, and pinching beamforming. The corresponding optimization problem is formulated as follows:
\begin{subequations}\label{Sum_Rate_Tri_Hybrid}
\allowdisplaybreaks[4]
\begin{align}
\max_{{\mathbf{W}}_{\rm{ana}},{\mathbf{w}}_{\rm{dig}},{\mathbf{W}}_{\rm{pin}}}&{\mathcal{R}}=\log_2(1+{\rm{SNR}})\\
{\rm{s.t.}}~\qquad&{\rm{tr}}({\mathbf{W}}_{\rm{ana}}{\mathbf{w}}_{\rm{dig}}{\mathbf{w}}_{\rm{dig}}^{H}{\mathbf{W}}_{\rm{ana}}^{H})\leq P,\label{Sum_Rate_Tri_Hybrid_Constraint1}\\
&\lvert[{\mathbf{W}}_{\rm{ana}}]_{m,\ell}\rvert=1,\forall m,\ell,\label{Sum_Rate_Tri_Hybrid_Constraint2}\\
&\lvert x_{m,n}-x_{m,n'}\rvert\geq\Delta_{\min},\forall m,n\ne n',\label{Sum_Rate_Tri_Hybrid_Constraint3}\\
& x_{m,n}\in[x_0,x_{\max}],\forall m,n,\label{Sum_Rate_Tri_Hybrid_Constraint4}
\end{align}
\end{subequations}
where $P>0$ is the power budget, $x_{\max}$ represents the maximum allowable PA deployment range, and $\Delta_{\min}$ denotes the minimum PA separation to avoid mutual coupling \cite{ivrlavc2010toward}.

The joint optimization of ${\mathbf{w}}_{\rm{dig}}$, ${\mathbf{W}}_{\rm{ana}}$, ${\mathbf{W}}_{\rm{pin}}$ is highly challenging due to the nonlinear coupling among the three beamforming components and the intricate dependence of both ${\mathbf{h}}$ and $\mathbf{G}$ on the PA positions. In the sequel, we develop an \emph{optimal tri-hybrid beamforming} design to derive analytical insights into the channel capacity of the proposed architecture.
\section{Tri-Hybrid Beamforming Design}
Problem \eqref{Sum_Rate_Tri_Hybrid} can be equivalently reformulated as follows:
\begin{subequations}\label{SU_Sum_Rate_Tri_Hybrid}
\allowdisplaybreaks[4]
\begin{align}
\max_{{\mathbf{W}}_{\rm{ana}},{\mathbf{w}}_{\rm{dig}},{\mathbf{W}}_{\rm{pin}}}&\lvert{\mathbf{h}}^{T}{\mathbf{G}}{\mathbf{W}}_{\rm{ana}}{\mathbf{w}}_{{\rm{dig}}}\rvert^2\\
{\rm{s.t.}}~\qquad&{\rm{tr}}({\mathbf{W}}_{\rm{ana}}{\mathbf{w}}_{\rm{dig}}{\mathbf{w}}_{\rm{dig}}^{H}{\mathbf{W}}_{\rm{ana}}^{H})\leq P,\\
&\eqref{Sum_Rate_Tri_Hybrid_Constraint2},\eqref{Sum_Rate_Tri_Hybrid_Constraint3},\eqref{Sum_Rate_Tri_Hybrid_Constraint4}.
\end{align}
\end{subequations}
For analytical tractability, we neglect the in-waveguide propagation loss by setting $\kappa=0$. This approximation is widely adopted in prior work \cite{ding2024flexible,ouyang2025array,wang2024antenna}, since its influence on the overall system performance is marginal compared with that of free-space path loss, which dominates in high-frequency bands. Accordingly, the results derived under this assumption represent the theoretical performance upper bound of the proposed tri-hybrid architecture. Under this idealized condition, it holds that $\alpha_{m,n}=\frac{1}{N}$, which means that the input power of each waveguide is uniformly distributed among its $N$ active PAs \cite{wang2025modeling}. This is a commonly used assumption in PASS \cite{ding2024flexible,ouyang2025array,wang2024antenna}, which simplifies the signal model and facilitates closed-form analysis in the subsequent derivations.
\subsection{Single-RF Scenario}
We commence with the single-RF case ($N_{\rm{rf}}=1$), in which the digital precoder ${\mathbf{w}}_{\rm{dig}}$ reduces to a scalar $w_{\rm{dig}}\in{\mathbb{C}}$, and the analog precoder ${\mathbf{W}}_{\rm{ana}}$ becomes a vector ${\mathbf{w}}_{\rm{ana}}\in{\mathbb{C}}^{M\times1}$. As a result, the power constraint simplifies as follows:
\begin{subequations}
\begin{align}
{\rm{tr}}({\mathbf{W}}_{\rm{ana}}{\mathbf{w}}_{\rm{dig}}{\mathbf{w}}_{\rm{dig}}^{H}{\mathbf{W}}_{\rm{ana}}^{H})&=\lvert{w_{\rm{dig}}}\rvert^2\lVert{{\mathbf{w}}_{\rm{ana}}}\rVert^2\\
&=M\lvert{w_{\rm{dig}}}\rvert^2\leq P.
\end{align} 
\end{subequations}
The objective function becomes
\begin{align}\label{SRF_SU_Power_Received}
\lvert{\mathbf{h}}^{T}{\mathbf{G}}{\mathbf{w}}_{\rm{ana}}\rvert^2\lvert{w_{\rm{dig}}}\rvert^2\leq
{P}\lvert{\mathbf{h}}^{T}{\mathbf{G}}{\mathbf{w}}_{\rm{ana}}\rvert^2/{M},
\end{align}
where equality is achieved by setting $w_{\rm{dig}}=\sqrt{{P}/{M}}$. Therefore, the optimal analog beamformer satisfies
\begin{align}
[{\mathbf{w}}_{\rm{ana}}]_m={\rm{e}}^{-{\rm{j}}\angle[{\mathbf{h}}^{T}{\mathbf{G}}]_m},\forall m\in{\mathcal{M}}.
\end{align}
Substituting this into \eqref{SU_Sum_Rate_Tri_Hybrid}, the optimization reduces to the maximization of the array gain as follows:
\begin{align}\label{SRF_SU_Sum_Rate_Tri_Hybrid}
\max\nolimits_{{\mathbf{W}}_{\rm{pin}}}~\sum\nolimits_{m=1}^{M}\lvert{\mathbf{h}}_m^T{\mathbf{g}}_m\rvert\quad{\rm{s.t.}}~\eqref{Sum_Rate_Tri_Hybrid_Constraint3},
\eqref{Sum_Rate_Tri_Hybrid_Constraint4},
\end{align}
where ${\mathbf{h}}_{m}\triangleq[h_{m,1},\ldots,h_{m,N}]^T$ denotes the channel vector from the $m$th waveguide to the user. To solve this, one must optimize the PA positions ${\mathbf{w}}_{{\rm{pin}},m}\triangleq[x_{m,1},\ldots,x_{m,N}]^T$ along each waveguide to maximize the array gain $\lvert{\mathbf{h}}_m^T{\mathbf{g}}_m\rvert$. This results in the following optimization:
\begin{subequations}\label{SRF_SU_Sum_Rate_Tri_Hybrid_Basic}
\allowdisplaybreaks[4]
\begin{align}
\max_{{\mathbf{w}}_{{\rm{pin}},m}}~&\lvert{\mathbf{h}}_m^T{\mathbf{g}}_m\rvert=\left\lvert\sum_{n=1}^{N}\frac{\sqrt{\eta}{\rm{e}}^{-{\rm{j}}\frac{2\pi}{\lambda}r_{m,n}-{\rm{j}}\frac{2\pi}{\lambda_{\rm{g}}}x_{m,n}}}
{\sqrt{N}r_{m,n}}\right\rvert\\
\quad{\rm{s.t.}}~&\eqref{Sum_Rate_Tri_Hybrid_Constraint3},\eqref{Sum_Rate_Tri_Hybrid_Constraint4}.
\end{align}
\end{subequations}
Despite being NP-hard, problem \eqref{SRF_SU_Sum_Rate_Tri_Hybrid_Basic} can be effectively addressed using a PA position refinement algorithm \cite{xu2024rate,ouyang2025array}. Without loss of generality, we assume $N$ is an even integer.
\subsubsection{Antenna Position Refinement}\label{Section: Single-RF Scenario: Antenna Position Refinement}
According to \cite{xu2024rate,ouyang2025array}, $\lvert{\mathbf{h}}_m^T{\mathbf{g}}_m\rvert$ is maximized when the received signals from all PAs are constructively combined at the user. This condition is achieved by aligning the total phase shift contributed by both free-space and in-waveguide propagation across all active PAs. For analytical tractability, the user is assumed to be located at the center of the service region. In this case, the user's SNR reaches its maximum when half of the PAs on each waveguide are positioned to the left of the user's horizontal coordinate $x_{\rm{u}}$, and the other half to the right \cite{xu2024rate}. Without loss of generality, we assume $x_{m,N}\leq\ldots\leq x_{m,\frac{N}{2}+1}\leq x_{\rm{u}}\leq x_{m,1}\leq\ldots\leq x_{m,\frac{N}{2}}$, and focus the analysis on $x_{m,n}$ for $n=1,\ldots,\frac{N}{2}$. 

For the $m$th waveguide, the initial position of the $(m,1)$th PA is set as $x_{m,1}=x_{\rm{u}}+\frac{\Delta_{\min}}{2}$ to satisfy the minimum inter-element spacing constraint. The PA is then shifted by a distance $v_{1}^{m}>0$ to the right in order to satisfy the phase alignment condition:
\begin{equation}\label{Refininement_Equation}
\sqrt{H_m^2+(\Delta_{1}^{m}+v_{1}^{m})^2}
+{(\Delta_{1}^{m}+v_{1}^{m})}n_{\rm{eff}}=H_{m,n},
\end{equation}
where $H_{m,1}=\lambda\lceil\frac{1}{\lambda}(\sqrt{H_m^2+(\Delta_{1}^{m})^2}+\Delta_{1}^{m}n_{\rm{eff}})\rceil$ and $\Delta_{1}^{m}=x_{m,1}-x_{\rm{u}}$. The solution is given by
\begin{align}\nonumber
v_{1}^{m,\star}=\left\{\begin{matrix}\frac{H_{m,1}n_{\rm{eff}}-\sqrt{H_{m,1}^2+H_m^2(n_{\rm{eff}}^2-1)}}{n_{\rm{eff}}^2-1}-\Delta_{n}^{m},&n_{\rm{eff}}\ne 1\\
\frac{H_{m,1}^2-H_m^2}{2H_{m,1}}-\Delta_{1}^{m},&n_{\rm{eff}}=1\end{matrix}\right..
\end{align}
Since a propagation distance of one wavelength results in a $2\pi$-phase shift and the left-hand side of \eqref{Refininement_Equation} increases monotonically with $v_1^m$, the optimal shift $v_{1}^{m,\star}$ is on the wavelength scale. Given that $\lambda\ll H_m$, the influence of this refinement on large-scale path loss is negligible.

After obtaining $v_{1}^{m,\star}$, the PA position is updated as $x_{1}^{m}\leftarrow x_{1}^{m}+v_{1}^{m,\star}$. The next PA is initialized as $x_{2}^{m}=x_{1}^{m}+\Delta_{\min}$ to maintain the minimum spacing, and its refinement $v_{2}^{m,\star}$ is computed in the same manner. Repeating this procedure iteratively yields the refined PA positions:
\begin{align}\label{Refined_Antenna_Location}
x_{m,n}^{\star}=x_{\rm{u}}+\left(n-1/2\right)\Delta_{\min}+\sum\nolimits_{i=1}^{n}v_{i}^{m,\star},~n\geq1.
\end{align}
This refinement ensures that the optimized PA locations satisfy the phase alignment condition:
\begin{equation}
\begin{split}
&(r_{m,1}+n_{\rm{eff}}x_{m,1})\mod \lambda\\
&=\ldots=(r_{m,{N}/{2}}+n_{\rm{eff}}x_{m,{N}/{2}})\mod \lambda.
\end{split}
\end{equation}
which guarantees constructive signal combination at the user. Additional implementation details for the general case, where the user is not located at the center, can be found in \cite{ouyang2025array,xu2024rate}. For analytical simplicity, the subsequent derivations focus on the centered-user scenario.
\subsubsection{Performance Upper and Lower Bounds}
To evaluate the performance of the antenna refinement method, we derive upper and lower bounds on the maximum achievable array gain. Subject to the minimum inter-element spacing constraint, the array gain admits the following upper bound 
\addtolength{\topmargin}{0.05in}
\cite{ouyang2025array}:
\begin{align}\label{Array_Gain_Fixed_Spacing_General2b}
\max_{{\mathbf{w}}_{{\rm{pin}},m}}~\lvert{\mathbf{h}}_m^T{\mathbf{g}}_m\rvert\leq\sum_{n=1}^{N/2}\frac{2\sqrt{\eta}}
{\sqrt{N}\sqrt{\left(n-1/2\right)^2\Delta_{\min}^2+H_{m}^2}}.
\end{align}
This bound is derived by ignoring phase terms and enforcing uniform spacing $\Delta_{\min}$. Equation \eqref{Refined_Antenna_Location} shows that the proposed refinement method closely approaches the upper bound in \eqref{Array_Gain_Fixed_Spacing_General2b} when $H_m\gg\sum\nolimits_{i=1}^{n}v_{i}^{m,\star}$, which is a mild condition given that each $v_{i}^{m,\star}$ lies within the wavelength scale and $H_m\gg \lambda$. Under such conditions, the maximum array gain is well approximated as follows: 
\begin{align}\label{Array_Gain_Basic_Summation}
\max_{{\mathbf{w}}_{{\rm{pin}},m}}~\lvert{\mathbf{h}}_m^T{\mathbf{g}}_m\rvert\approx \sum_{n=1}^{N/2}\frac{2\sqrt{\eta}}
{\sqrt{N}\sqrt{\left(n-1/2\right)^2\Delta_{\min}^2+H_{m}^2}}.
\end{align}

For completeness, we also derive a lower bound on the maximum array gain. When the gain is maximized, phase alignment across all antennas is achieved, which yields
\begin{align}\label{Array_Gain_Maximum}
\max_{{\mathbf{w}}_{{\rm{pin}},m}}~\lvert{\mathbf{h}}_m^T{\mathbf{g}}_m\rvert=\sum\nolimits_{n=1}^{N}\frac{\sqrt{\eta}}
{\sqrt{N}\sqrt{(x_{m,n}^{\star}-x_{\rm{u}})^2+H_{m}^2}}.
\end{align}
If $\Delta_{\max}$ represents the largest inter-antenna spacing among these positions, then the gain is lower bounded by
\begin{align}\label{Array_Gain_Maximum_Lower_Bound}
\max_{{\mathbf{w}}_{{\rm{pin}},m}}~\lvert{\mathbf{h}}_m^T{\mathbf{g}}_m\rvert\geq\sum_{n=1}^{N/2}\frac{2\sqrt{\eta}}
{\sqrt{N}\sqrt{\left(n-1/2\right)^2\Delta_{\max}^2+H_{m}^2}}.
\end{align}
This result is derived by enforcing uniform spacing $\Delta_{\max}$ in \eqref{Array_Gain_Maximum}. According to the design of the antenna position refinement algorithm, $\Delta_{\max}$ is on the order of the wavelength.

Using these bounds, we next characterize the power scaling behavior of the tri-hybrid beamforming architecture as a function of the number of antennas $N$.
\subsubsection{Power Scaling Law}
In practical systems, $\Delta_{\min}$ is typically on the order of the wavelength, which satisfies $\Delta_{\min}\ll H_m$ \cite{ivrlavc2010toward,xu2024rate}. Under this condition, the right-hand side of \eqref{Array_Gain_Basic_Summation} can be approximated as follows.
\vspace{-5pt}
\begin{lemma}\label{Lemma_Power_Scaling_Law}
Given that $\Delta_{\min}\ll H_m$, the maximum array gain for the $m$th waveguide can be approximated as follows:
\begin{align}\label{Lemma_Power_Scaling_Law_Main}
\max_{{\mathbf{w}}_{{\rm{pin}},m}}~\lvert{\mathbf{h}}_m^T{\mathbf{g}}_m\rvert\approx \frac{2\sqrt{\eta}}{\sqrt{N}\Delta_{\min}}f_{\rm{ub}}\left(\frac{N\Delta_{\min}}{2H_m}\right).
\end{align}
where $f_{\rm{ub}}(x)\triangleq\ln(\sqrt{1+x^2}+x)$.
\end{lemma}
\vspace{-5pt}
\begin{IEEEproof}
The right-hand side of \eqref{Array_Gain_Basic_Summation} satisfies
\begin{align}\label{Array_Gain_Summation_Integral}
\frac{2\sqrt{\eta}}{\sqrt{N}\Delta_{\min}}\sum\nolimits_{n=1}^{N/2}\frac{\Delta_{\min}}{H_{m}}\frac{1}
{\sqrt{\left(n-1/2\right)^2(\frac{\Delta_{\min}}{H_{m}})^2+1}}.
\end{align}
Assuming $\frac{\Delta_{\min}}{H_{m}}\ll1$, we approximate the sum using the concept of definite integrals. Letting $f_{\rm{int}}(x)\triangleq\frac{1}{\sqrt{x^2+1}}$ gives
\begin{subequations}
\allowdisplaybreaks[4]
\begin{align}
\eqref{Array_Gain_Summation_Integral}&=\frac{2\sqrt{\eta}}{\sqrt{N}\Delta_{\min}}\sum_{n=1}^{N/2}\frac{\Delta_{\min}}{H_{m}}
f_{\rm{int}}\left(\left(n-\frac{1}{2}\right)\frac{\Delta_{\min}}{H_{m}}\right)\\
&=\frac{2\sqrt{\eta}}{\sqrt{N}\Delta_{\min}}\int_{0}^{\frac{N\Delta_{\min}}{2H_m}}
f_{\rm{int}}\left(x\right){\rm{d}}x,
\end{align}
\end{subequations}
which, together with \cite[Eq. (2.01.18)]{gradshteyn2014table}, leads to the result in \eqref{Lemma_Power_Scaling_Law_Main}. This concludes the proof.
\end{IEEEproof}
Combining \eqref{Lemma_Power_Scaling_Law_Main} with \eqref{SRF_SU_Sum_Rate_Tri_Hybrid}, we approximate the received SNR in the single-RF case as follows:
\begin{subequations}\label{Upper_Bound_Appr}
\allowdisplaybreaks[4]
\begin{align}
{\rm{SNR}}_{1}&=\frac{P}{M\sigma^2}\left(\max_{{\mathbf{W}}_{\rm{pin}}}~\sum\nolimits_{m=1}^{M}\lvert{\mathbf{h}}_m^T{\mathbf{g}}_m\rvert\right)^2\\
&\approx\frac{P}{M\sigma^2}\frac{4{\eta}}{{N}\Delta_{\min}^2}\left(\sum\nolimits_{m=1}^{M}f_{\rm{ub}}\left(\frac{N\Delta_{\min}}{2H_m}\right)\right)^2.\label{Approximation_Upper_Bound}
\end{align}
\end{subequations}
Given that $\frac{\Delta_{\min}}{H_m}\ll1$, we typically have $\frac{N\Delta_{\min}}{2H_m}\ll1$ for small to moderately large values of $N$. Applying the first-order Taylor expansion $f_{\rm{ub}}(x)\approx x$ as $x\rightarrow0$, we obtain the following power scaling law:
\begin{align}\label{Linear_SNR}
{\rm{SNR}}_1\approx \frac{PN\eta}{M\sigma^2}\left(\sum\nolimits_{m=1}^{M}\frac{1}{H_m}\right)^2.
\end{align}
On this basis, we make the following observation.
\vspace{-5pt}
\begin{remark}\label{Remark_Linear_SNR}
When $N\Delta_{\min}\ll H_m$, the received SNR scales linearly with the number of pinching antennas per waveguide. Besides, the linear scaling rate is not impacted by $\Delta_{\min}$.
\end{remark}
\vspace{-5pt}
Next, we consider the asymptotic regime $N\rightarrow\infty$, where \eqref{Lemma_Power_Scaling_Law_Main} no longer yields a precise approximation of the array gain. To handle this difficulty, we rely instead on the upper and lower bounds to analyze the power scaling behavior.
\vspace{-5pt}
\begin{lemma}\label{Lemma_Power_Scaling_Law_Real}
When $N\rightarrow\infty$, the received SNR scales as ${\mathcal{O}}((\ln{N})^2/N)$ with the number of PAs.
\end{lemma}
\vspace{-5pt}
\begin{IEEEproof}
As previously discussed, $\Delta_{\max}$ is on the order of the wavelength, which satisfies $\Delta_{\max}\ll H_m$. Under this condition, the right-hand side of \eqref{Array_Gain_Maximum_Lower_Bound}, i.e., the lower bound on the array gain, can be approximated as $ \frac{2\sqrt{\eta}}{\sqrt{N}\Delta_{\max}}f_{\rm{ub}}\left(\frac{N\Delta_{\max}}{2H_m}\right)$. Accordingly, the lower bound on the received SNR satisfies
\begin{align}\label{Lower_Bound_Appr}
{\rm{SNR}}_{1}\geq\frac{P}{M\sigma^2}\frac{4{\eta}}{{N}\Delta_{\max}^2}\left(\sum\nolimits_{m=1}^{M}f_{\rm{ub}}\left(\frac{N\Delta_{\max}}{2H_m}\right)\right)^2.
\end{align}
On the other hand, according to \eqref{Array_Gain_Fixed_Spacing_General2b} and \eqref{Lemma_Power_Scaling_Law_Main}, the SNR is upper bounded by \eqref{Approximation_Upper_Bound}. Due to the logarithmic growth of $f_{\rm{ub}}(x)$ with respect to $x$, both the upper and lower bounds on SNR scale as ${\mathcal{O}}((\ln{N})^2/N)$ in the limit as $N\rightarrow\infty$. By applying the squeeze theorem, we conclude $\lim_{N\rightarrow\infty}{\rm{SNR}}_1\simeq{\mathcal{O}}((\ln{N})^2/N)$. This concludes the proof.
\end{IEEEproof}
This implies that although increasing $N$ initially improves the SNR, the benefit diminishes with large $N$, and ultimately $\lim_{N\rightarrow\infty}{\rm{SNR}}_1=\lim_{N\rightarrow\infty}\frac{(\ln{N})^2}{N}=0$.
\vspace{-5pt}
\begin{remark}\label{Remark_2_Lim}
The above result can be interpreted as follows. As $N\rightarrow\infty$, the power per PA, i.e., $\frac{P}{MN}$, decreases. Therefore, the PAs that account for the majority of the power are positioned too far from the user, which makes the user receive negligible energy from the PAs.
\end{remark}
\vspace{-5pt}
By combining \eqref{Array_Gain_Fixed_Spacing_General2b} with \eqref{Array_Gain_Maximum_Lower_Bound}, as well as \eqref{Upper_Bound_Appr} and \eqref{Lower_Bound_Appr}, the following upper and lower bounds on the channel capacity ${\mathcal{R}}_{1}\triangleq\log_2(1+{\rm{SNR}}_1)$ can be derived as follows:
\begin{align}\label{Bound_Capacity_1}
&\log_2\left(1+\frac{\frac{P}{M\sigma^2}4{\eta}}{{N}\Delta_{\max}^2}\left(\sum\nolimits_{m=1}^{M}f_{\rm{ub}}\left(\frac{N\Delta_{\max}}{2H_m}\right)\right)^2\right)\leq{\mathcal{R}}_1
\nonumber\\&\leq
\log_2\left(1+\frac{\frac{P}{M\sigma^2}4{\eta}}{{N}\Delta_{\min}^2}\left(\sum\nolimits_{m=1}^{M}f_{\rm{ub}}\left(\frac{N\Delta_{\min}}{2H_m}\right)\right)^2\right).
\end{align}
From \eqref{Bound_Capacity_1}, it follows that the channel capacity scales with the number of PAs $N$ as follows:
\begin{align}
\lim\nolimits_{N\rightarrow\infty}{\mathcal{R}}_1\simeq{\mathcal{O}}\left(\log_2\left({(\ln{N})^2}/{N}\right)\right).
\end{align}
Therefore, the capacity satisfies $\lim_{N\rightarrow\infty}{\mathcal{R}}_1=0$, which is consistent with the observation in Remark \ref{Remark_2_Lim}.
\subsection{Multiple-RF Scenario}
We now extend the analysis to the multiple-RF scenario ($N_{\rm{rf}}>1$), where the digital and analog precoders ${\mathbf{w}}_{{\rm{dig}}}$ and ${\mathbf{W}}_{\rm{ana}}$ can be jointly designed to align the transmitted signal with the effective channel vector ${\mathbf{h}}^{T}{\mathbf{G}}$. That is, the optimal beamformer satisfies ${\mathbf{W}}_{\rm{ana}}{\mathbf{w}}_{{\rm{dig}}}=({\mathbf{h}}^{T}{\mathbf{G}})^{H}$, which maximizes the received SNR \cite{zhang2005variable}. Notably, this beam alignment can be realized using only $N_{\rm{rf}}=2$ RF chains \cite{zhang2005variable}. Under this design, the received SNR is given by
\begin{align}\label{MRF_SU_Power_Received}
{\rm{SNR}}_2=\lvert{\mathbf{h}}^{T}{\mathbf{G}}{\mathbf{W}}_{\rm{ana}}{\mathbf{w}}_{{\rm{dig}}}\rvert^2/{\sigma^2}
={P}\lVert{\mathbf{h}}^{T}{\mathbf{G}}\rVert^2/{\sigma^2}.
\end{align}
Substituting this into \eqref{SU_Sum_Rate_Tri_Hybrid}, the optimization reduces to
\begin{align}\label{MRF_SU_Sum_Rate_Tri_Hybrid}
\max_{{\mathbf{W}}_{\rm{pin}}}~\sum\nolimits_{m=1}^{M}\lvert{\mathbf{h}}_m^T{\mathbf{g}}_m\rvert^2\quad{\rm{s.t.}}~\eqref{Sum_Rate_Tri_Hybrid_Constraint3},
\end{align} 
which can be addressed using the antenna refinement strategy described in Section \ref{Section: Single-RF Scenario: Antenna Position Refinement}. Following the derivations used to establish Lemma \ref{Lemma_Power_Scaling_Law}, the corresponding SNR can be approximated as follows:
\begin{subequations}
\allowdisplaybreaks[4]
\begin{align}
{\rm{SNR}}_2&=\frac{P}{\sigma^2}\left(\max_{{\mathbf{W}}_{\rm{pin}}}~\sum\nolimits_{m=1}^{M}\lvert{\mathbf{h}}_m^T{\mathbf{g}}_m\rvert^2\right)\\
&\approx\frac{P}{\sigma^2}\frac{4{\eta}}{{N}\Delta_{\min}^2}\sum\nolimits_{m=1}^{M}\left(f_{\rm{ub}}\left(\frac{N\Delta_{\min}}{2H_m}\right)\right)^2.
\end{align}
\end{subequations}
Applying the Cauchy-Schwarz inequality, we note that ${\rm{SNR}}_1\leq{\rm{SNR}}_2$. By following a similar approach to the power scaling analysis in the single-RF case, we characterize the scaling behavior in the multiple-RF setting. In the regime $\frac{N\Delta_{\min}}{2H_m}\ll1$, the scaling behavior becomes
\begin{align}
{\rm{SNR}}_2\approx \frac{PN\eta}{\sigma^2}\left(\sum\nolimits_{m=1}^{M}\frac{1}{H_m^2}\right).
\end{align} 
The capacity ${\mathcal{R}}_{2}\triangleq\log_2(1+{\rm{SNR}}_2)$ is bounded as follows:
\begin{align}
&\log_2\left(1+\frac{\frac{P}{M\sigma^2}4{\eta}}{{N}\Delta_{\max}^2}\sum\nolimits_{m=1}^{M}\left(f_{\rm{ub}}\left(\frac{N\Delta_{\max}}{2H_m}\right)\right)^2\right)\leq{\mathcal{R}}_2
\nonumber\\&\leq
\log_2\left(1+\frac{\frac{P}{M\sigma^2}4{\eta}}{{N}\Delta_{\min}^2}\sum\nolimits_{m=1}^{M}\left(f_{\rm{ub}}\left(\frac{N\Delta_{\min}}{2H_m}\right)\right)^2\right).\label{Bound_Capacity_2}
\end{align}
As $N\rightarrow\infty$, we observe
\begin{subequations}
\begin{align}
{\rm{SNR}}_2&\simeq{\mathcal{O}}((\ln{N})^2/N),\\
{\mathcal{R}}_2&\simeq{\mathcal{O}}\left(\log_2\left({(\ln{N})^2}/{N}\right)\right).
\end{align}
\end{subequations}
Comparing ${\rm{SNR}}_2$ with ${\rm{SNR}}_1$ yields the following result. 
\vspace{-5pt}
\begin{remark}
${\rm{SNR}}_2$ and ${\rm{SNR}}_1$ exhibit similar trends: they both increase approximately linearly with $N$ for small $N$, and eventually decrease as $N$ grows large. This highlights that simply increasing the number of pinching antennas does not lead to continuous SNR improvement. An optimal antenna number exists for both single- and multiple-RF chain configurations.
\end{remark}
\vspace{-5pt}
\subsection{Extension}
In the above discussion, the performance of a single user under a TDMA framework was considered. It was assumed that pinching beamforming is redesigned in each time slot for the scheduled user so that the signals radiated by different PAs combine constructively at the intended receiver. This strategy achieves the performance upper bound, but it incurs relatively high implementation cost, since the PA locations must be adjusted in every time slot according to the scheduled user's location. This operating mode is referred to as \emph{pinching switching (PS)} \cite{liu2026survey}. An alternative approach that reduces implementation cost is \emph{pinching multiplexing (PM)}, where all users share a common pinching beamformer across different time slots, i.e., the PA placement remains fixed. Under this scheme, the achievable user rate is generally reduced, but the system implementation becomes significantly simpler because PA repositioning across time slots is avoided.

For simplicity, it is assumed that the digital and analog beamformers of each user are redesigned within its dedicated time slot. These beamformers can be updated electronically with relatively low overhead, whereas adjusting PA positions may require mechanical operations and therefore incurs higher latency and complexity. Under this assumption, the achievable sum-rate of PM-based TDMA can be formulated as follows:
\begin{align}
{\mathcal{R}}_{\rm{sum}}\triangleq\frac{1}{K}\sum\nolimits_{k=1}^{K}\log_2(1+{P}\lVert\hat{\mathbf{h}}_k^{T}{\mathbf{G}}\rVert^2/{\sigma^2}),
\end{align}
where $P$ denotes the per-user transmit power budget, $\sigma^2$ denotes the noise power, and $\hat{\mathbf{h}}_k\in{\mathbb{C}}^{MN\times1}$ represents the channel vector from all pinched waveguides to user $k$. Following \eqref{SRF_SU_Sum_Rate_Tri_Hybrid_Basic}, the effective channel gain $a_k({\mathbf{W}}_{\rm{pin}})\triangleq\lVert\hat{\mathbf{h}}_k^{T}{\mathbf{G}}\rVert^2$ for each user $k$ can be written as follows:
\begin{align}
a_k({\mathbf{W}}_{\rm{pin}})=\sum_{m=1}^{M}\left\lvert\sum_{n=1}^{N}\frac{\sqrt{\eta}{\rm{e}}^{-{\rm{j}}\frac{2\pi}{\lambda}r_{m,n}^{(k)}-{\rm{j}}\frac{2\pi}{\lambda_{\rm{g}}}x_{m,n}}}
{\sqrt{N}r_{m,n}^{(k)}}\right\rvert ,
\end{align}
where $[{\mathbf{W}}_{\rm{pin}}]_{m,n}=x_{m,n}$ for $m\in{\mathcal{M}}$ and $n\in{\mathcal{N}}$. The position of user $k$ is denoted by ${\mathbf{r}}_k=[x_{{\rm{u}}}^{(k)},y_{{\rm{u}}}^{(k)},0]^T\in{\mathbb{R}}^{3\times1}$, and $r_{m,n}^{(k)}\triangleq\lVert{\bm\psi}_{m,n}-{\mathbf{r}}_k\rVert=\sqrt{(x_{m,n}-x_{{\rm{u}}}^{(k)})^2+H_{k,m}^2}$, with $H_{k,m}\triangleq\sqrt{(y_m-y_{{\rm{u}}}^{(k)})^2+H^2}$ representing the effective elevation distance between user $k$ and the $m$th waveguide. Accordingly, the sum-rate maximization problem under PM-based TDMA can be formulated as follows:
\begin{subequations}\label{Sum_Rate_Tri_Hybrid_PM_TDMA}
\begin{align}
\max_{{\mathbf{W}}_{\rm{pin}}}~&{\mathcal{R}}_{\rm{sum}}=\sum\nolimits_{k=1}^{K}\log_2(1+{P}a_k({\mathbf{W}}_{\rm{pin}})/{\sigma^2})\\
{\rm{s.t.}}~&\eqref{Sum_Rate_Tri_Hybrid_Constraint3},\eqref{Sum_Rate_Tri_Hybrid_Constraint4}.
\end{align}
\end{subequations}
Finding the optimal pinching beamformer for PM is more challenging than for PS, since all users share the same pinching beamformer ${\mathbf{W}}_{\rm{pin}}$. As a result, ${\mathbf{W}}_{\rm{pin}}$ must be designed such that the signals intended for different users can be constructively combined at their respective receivers across different time slots, which generally leads to a difficult nonconvex optimization problem. As a compromise between performance and complexity, we employ an element-wise alternating optimization framework to efficiently obtain a high-quality suboptimal solution to the non-convex problem in \eqref{Sum_Rate_Tri_Hybrid_PM_TDMA}.

In the proposed approach, each antenna position $x_{m,n}$ is optimized sequentially while keeping all other coordinates fixed. Let ${\mathcal{R}}_{m,n}(x)$ denotes the sum-rate when $x_{m,n}$ in ${\mathbf{W}}_{\rm{pin}}$ is set to $x$ while others are fixed. This subproblem is equivalent to the following:
\begin{align}\label{MP_Pareto_Sub_sub1}
\max\nolimits_{x}~{\mathcal{R}}_{m,n}(x)~~{\rm{s.t.}}~\eqref{Sum_Rate_Tri_Hybrid_Constraint3},\eqref{Sum_Rate_Tri_Hybrid_Constraint4}.
\end{align}
Since this is a single-variable optimization over a bounded interval, it can be solved efficiently via one-dimensional search. To do so, we discretize the interval $[x_0,x_{\max}]$ into a uniform grid of $X$ points:
\begin{align}
{\mathcal{X}}\triangleq \left\{x_0,x_0+\frac{I_{\rm{P}}}{X-1},x_0+\frac{2I_{\rm{P}}}{X-1},\ldots,x_{\max}\right\},
\end{align}
where $I_{\rm{P}}\triangleq x_{\max}-x_0$. A near-optimal $x_{m,n}$, denoted as $x_{m,n}^{\star}$, is then selected according to the following:
\begin{align}
x_{m,n}^{\star}\triangleq\argmax\nolimits_{x_{m,n}\in {\mathcal{X}}\setminus {\mathcal{X}}_{m,n}}{\mathcal{R}}_{m,n}(x),
\end{align}
where the set ${\mathcal{X}}_{m,n}$ includes all grid points that violate the minimum spacing constraint, i.e.,
\begin{align}
{\mathcal{X}}_{m,n}\triangleq \{x|x\in{\mathcal{X}},\lvert x - x_{m,n'}\rvert<\Delta,n'\ne n\}.
\end{align}
This procedure is applied iteratively to all antennas until convergence. The complete algorithm is summarized in Algorithm \ref{Algorithm1}, with a computational complexity of ${\mathcal{O}}(I_{\rm{iter}} NMX)$, where $I_{\rm{iter}}$ is the number of iterations to convergence.

\begin{algorithm}[htbp] 
\algsetup{linenosize=\tiny} \scriptsize 
\caption{Element-wise Algorithm for Solving \eqref{Sum_Rate_Tri_Hybrid_PM_TDMA}}
\label{Algorithm1}
\begin{algorithmic}[1]
\STATE initialize the optimization variables
\REPEAT 
  \FOR{$n\in{\mathcal{N}}$ and $m\in{\mathcal{M}}$}
      \STATE update $x_{m,n}$ through one-dimensional search
    \ENDFOR
\UNTIL{convergence}
\end{algorithmic}
\end{algorithm}
\section{Numerical Results}
We perform numerical simulations to validate the analytical results. Unless stated otherwise, the system parameters are set as follows \cite{ouyang2025array}: carrier frequency $f_{\rm{c}} = 28$ GHz, effective refractive index $n_{\rm{eff}} = 1.4$, in-waveguide attenuation factor $\kappa = 0.08$ dB/m, minimum inter-PA spacing $\Delta_{\min}=\frac{\lambda}{2}$, power budget $P = 10$ dBm, and noise power $\sigma^2=-90$ dBm. The users are assumed to be uniformly distributed within a rectangular region centered at the origin, with side lengths $D_x=50$ m and $D_y = 20$ m along the $x$- and $y$-axes, respectively. The waveguides are aligned parallel to the $x$-axis at a fixed height $H = 3$ m. Their $y$-coordinates are given by $y_m=-\frac{D_y}{2}+\frac{m-1}{M-1}D_y$ for $m\in{\mathcal{M}}$ so that the waveguides uniformly span the entire horizontal coverage area. The $x$-coordinates of the feed points for all waveguides are set to $x_0=-\frac{D_x}{2}$.

\begin{figure}[!t]
\centering
\includegraphics[height=0.16\textwidth]{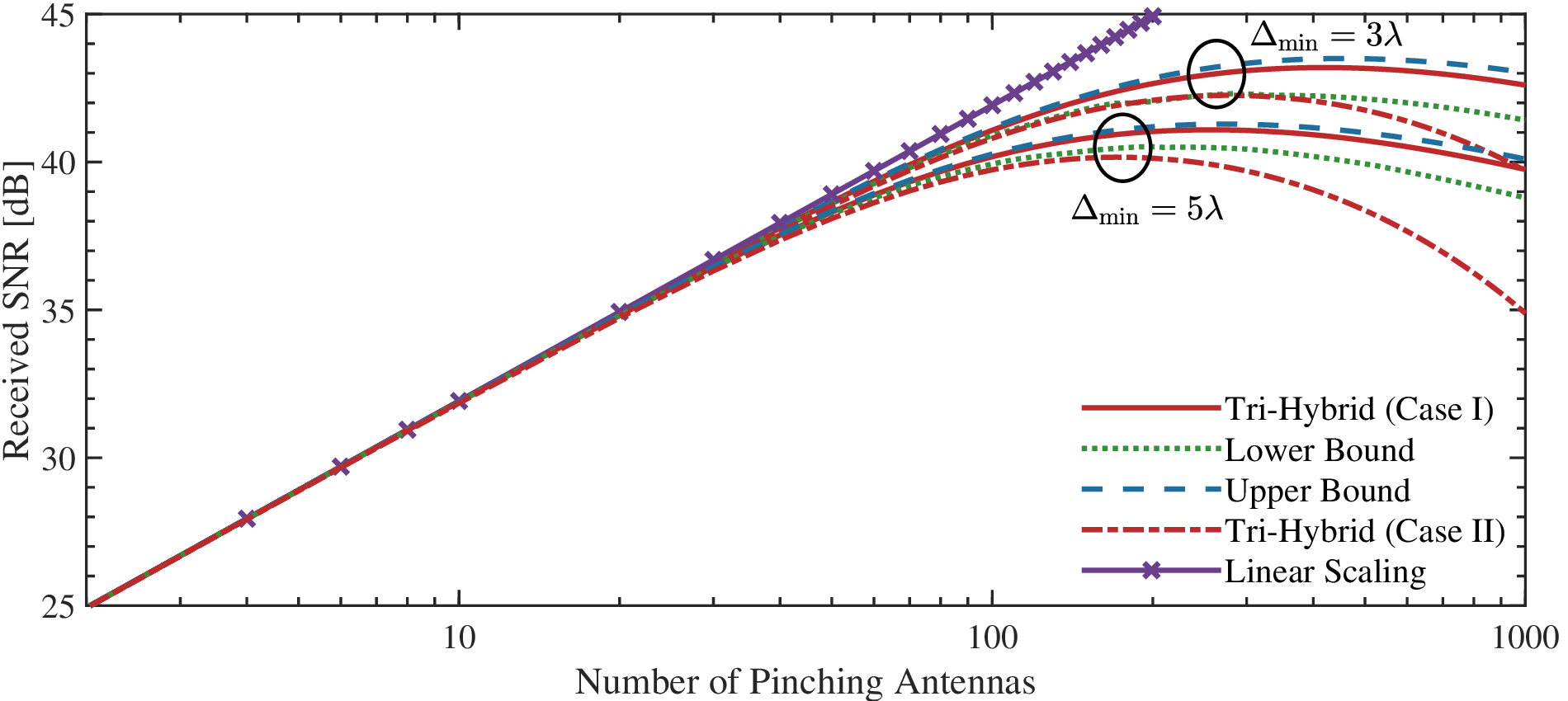}
\caption{Single-RF received SNR vs the number of PAs $N$. $x_{\rm{u}}=y_{\rm{u}}=0$ m.}
\label{Figure_SU_Array_Gain_Antenna_Number}
\vspace{-10pt}
\end{figure}

\begin{figure}[!t]
\centering
\includegraphics[height=0.16\textwidth]{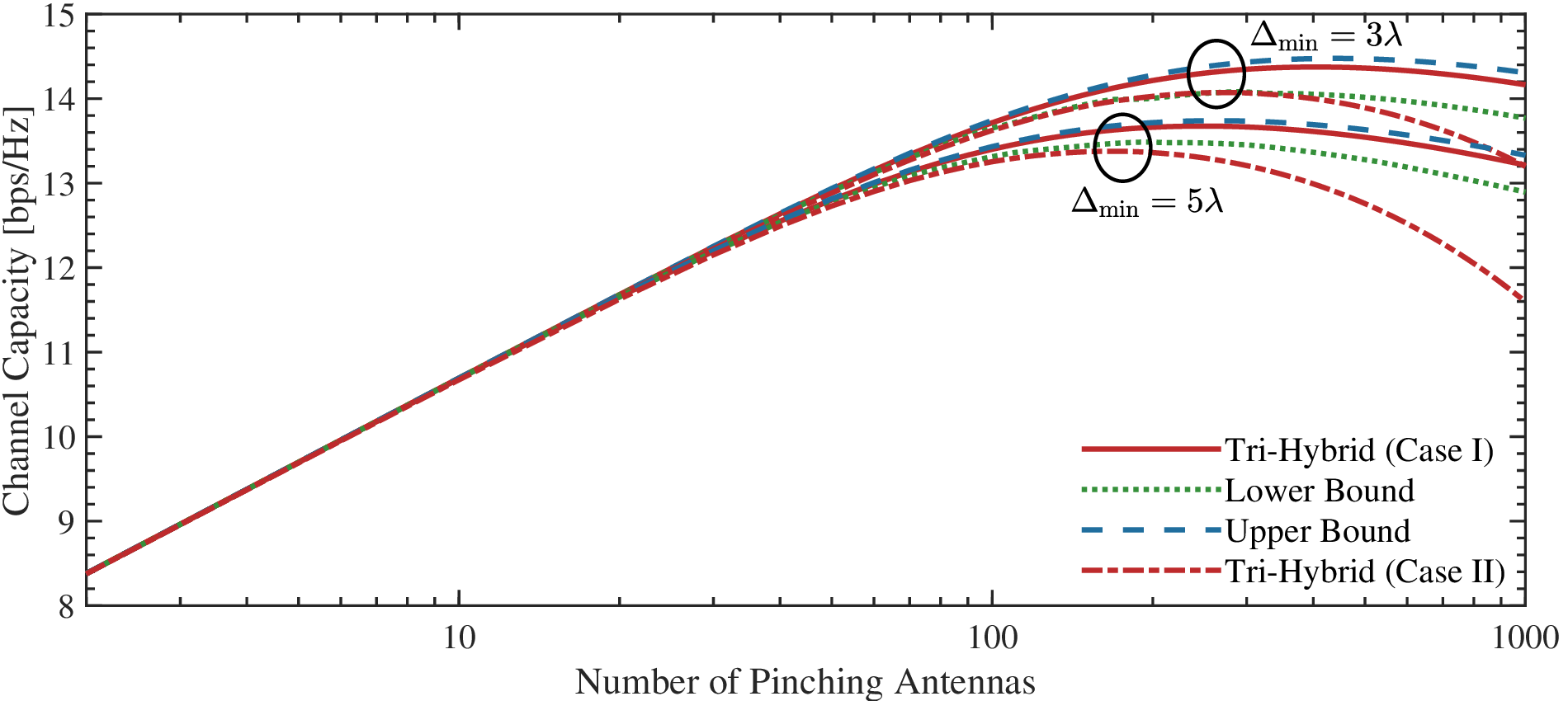}
\caption{Multiple-RF capacity vs the number of PAs $N$. $x_{\rm{u}}=y_{\rm{u}}=0$ m.}
\label{Figure_SU_Channel_Capacity_Antenna_Number}
\vspace{-10pt}
\end{figure}

{\figurename} {\ref{Figure_SU_Array_Gain_Antenna_Number}} illustrates the received SNR in the single-RF-chain case, i.e., ${\rm{SNR}}_1$, as a function of the number of PAs $N$ for different values of $\Delta_{\min}$. For completeness, two scenarios are presented: Case I, without in-waveguide propagation loss, and Case II, with in-waveguide loss included. In both cases, ${\rm{SNR}}_1$ increases approximately linearly with $N$ when the number of PAs is small. The simulated results agree well with the analytical linear asymptotic expression in \eqref{Linear_SNR}, which supports the conclusion stated in Remark \ref{Remark_Linear_SNR}. The upper (given by \eqref{Upper_Bound_Appr}) and lower bounds (given by \eqref{Lower_Bound_Appr}) of ${\rm{SNR}}_1$ are also shown. Both bounds reveal the same scaling trend with respect to $N$ as ${\rm{SNR}}_1$: first increasing, reaching a maximum, and then decreasing as $N$ continues to grow. This trend yields the existence of an optimal number of PAs that maximizes the SNR, as shown in {\figurename} {\ref{Figure_SU_Array_Gain_Antenna_Number}}. This observation aligns with the insight discussed in Remark \ref{Remark_2_Lim}. The influence of in-waveguide propagation loss remains negligible when $N$ is small. However, as $N$ becomes large, the accumulated attenuation inside the waveguide causes a noticeable SNR reduction in Case II compared with Case I. {\figurename} {\ref{Figure_SU_Channel_Capacity_Antenna_Number}} shows the channel capacity in the multiple-RF-chain case with $N_{\rm{rf}}=2$, i.e., ${\mathcal{R}}_2$, as a function of $N$. As observed, the capacity remains tightly confined within the analytical upper and lower bounds derived in \eqref{Bound_Capacity_2}. Moreover, an optimal number of PAs is observed to exist that maximizes the achievable channel capacity.

\begin{figure}[!t]
\centering
    \subfigure[$M=4$.]
    {
        \includegraphics[height=0.16\textwidth]{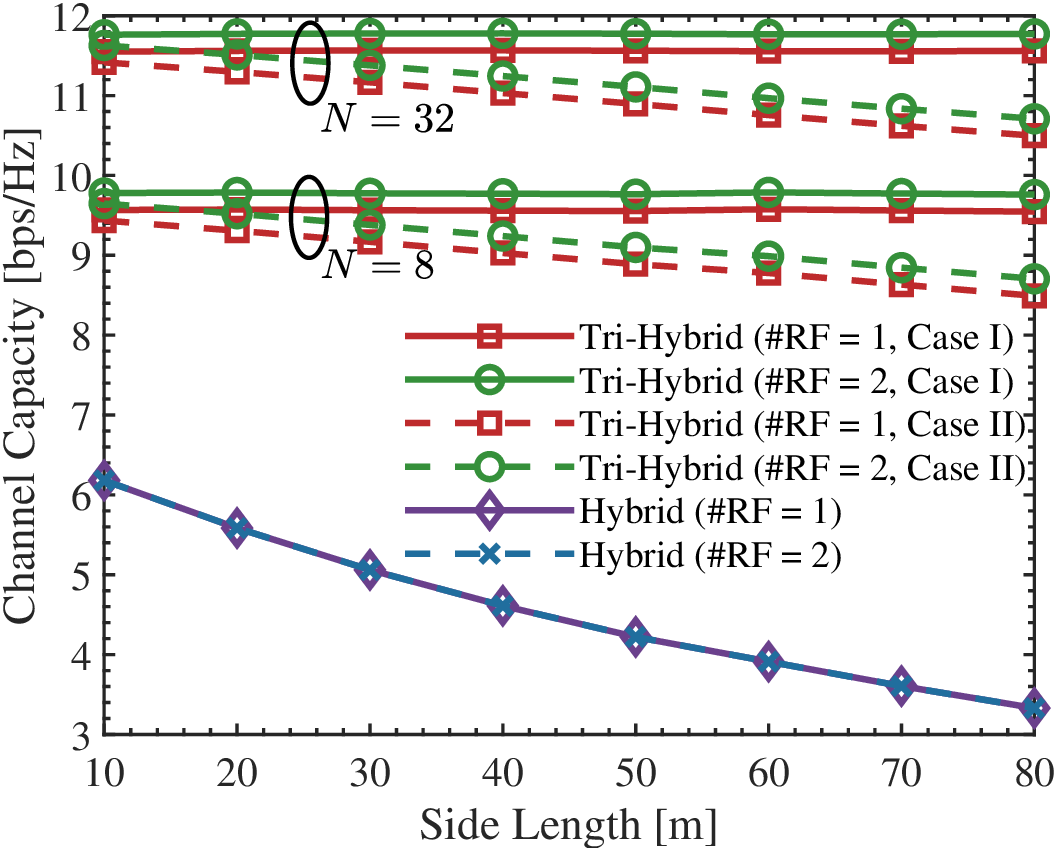}
	   \label{Figure_SU_Array_Gain_Side_Length}
    }
   \subfigure[$D_x=50$ m.]
    {
        \includegraphics[height=0.16\textwidth]{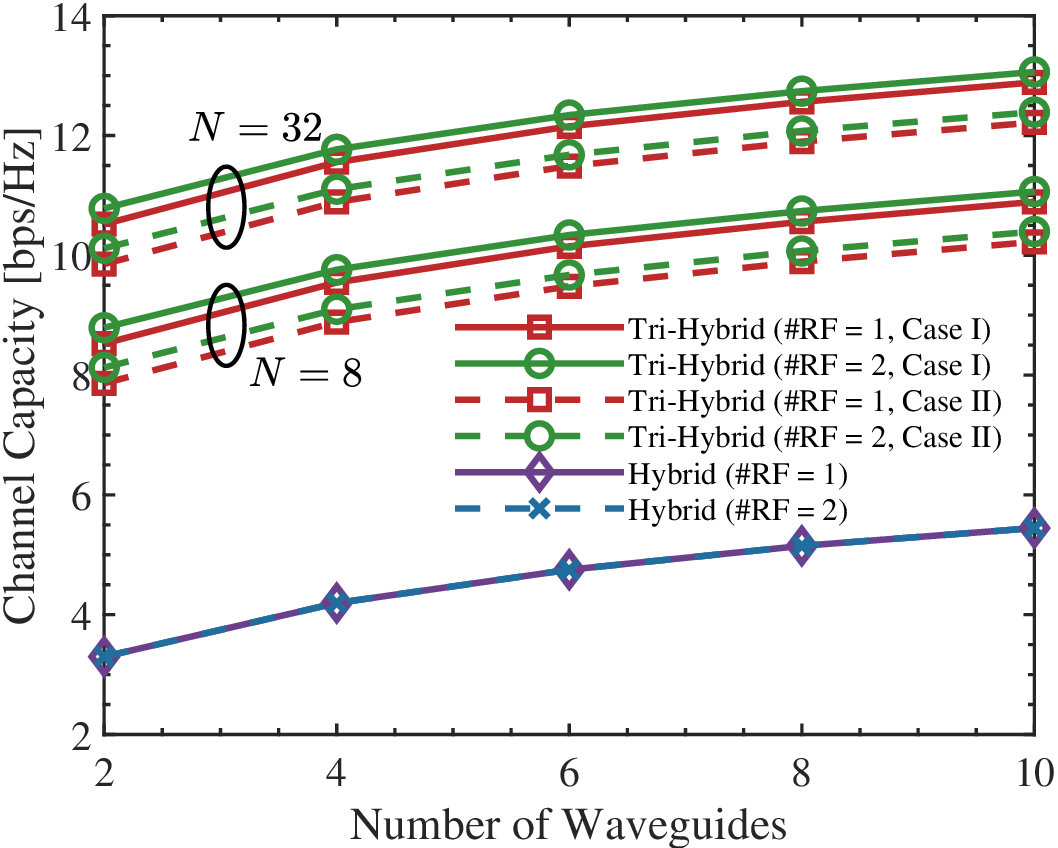}
	   \label{Figure_SU_Array_Gain_Waveguide}
    }
\caption{Average channel capacity.}
\label{Figure: Average capacity}
\vspace{-15pt}
\end{figure}

{\figurename} {\ref{Figure: Average capacity}} presents the average channel capacity of a user uniformly distributed within the service region. {\figurename} {\ref{Figure_SU_Array_Gain_Side_Length}} depicts the capacity as a function of the side length $D_x$. To highlight the advantages of the proposed tri-hybrid beamforming architecture, its performance is compared with that of a conventional hybrid digital-analog beamforming system without PAs. In the baseline configuration, a fixed hybrid array is placed at the center of the service region, using the same number of RF chains and phase shifters as the proposed design. As shown in {\figurename} {\ref{Figure_SU_Array_Gain_Side_Length}}, for both cases with and without in-waveguide propagation loss, the tri-hybrid beamforming achieves a significantly higher channel capacity than the conventional hybrid counterpart. This gain arises from the ability of the PAs to reconfigure the channel geometry in real time, which enables the establishment of stronger LoS links for individual users. Under the considered setup, the use of multiple RF chains provides only a modest gain over the single-RF-chain configuration, since only one data stream is transmitted. As $D_x$ increases, the average user-to-array distance also increases, causing higher path loss and a corresponding capacity reduction in the conventional hybrid scheme. In contrast, the tri-hybrid architecture adaptively positions the PAs closer to the users, maintaining a shorter transmission distance. As a result, the capacity remains nearly constant in Case I and decreases only slightly in Case II as the region expands. {\figurename} {\ref{Figure_SU_Array_Gain_Waveguide}} illustrates the capacity as a function of the number of waveguides $M$. As expected, both tri-hybrid and hybrid beamforming benefit from a larger $M$. However, the tri-hybrid architecture consistently outperforms the conventional scheme for all configurations, confirming its effectiveness in improving overall system throughput through dynamic spatial reconfiguration.

\begin{figure}[!t]
\centering
\includegraphics[height=0.16\textwidth]{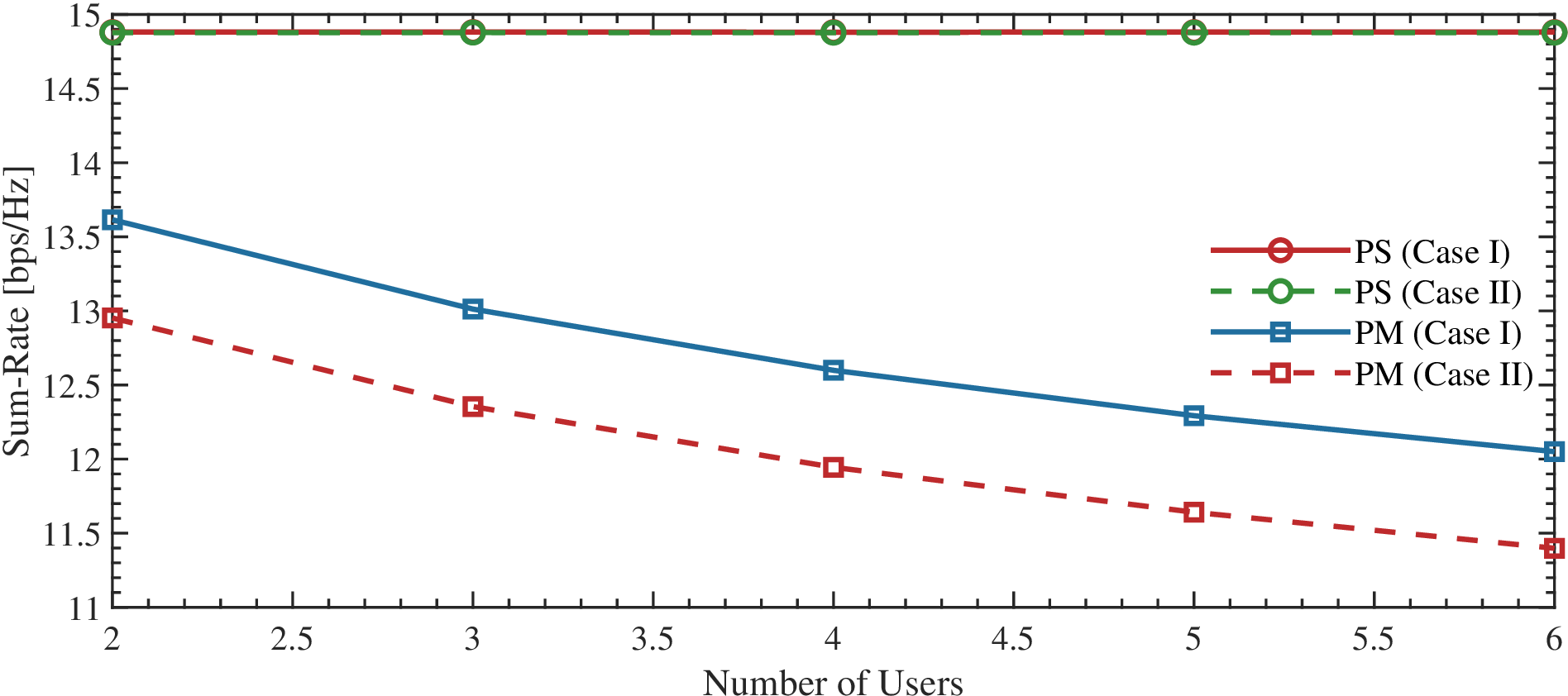}
\caption{Sum-rate versus the user number. $M=4$ and $N=32$.}
\label{Figure_Multiple_User_TDMA_User_Number}
\vspace{-15pt}
\end{figure}

{\figurename} {\ref{Figure_Multiple_User_TDMA_User_Number}} compares the sum-rate achieved by PS and PM as a function of the number of users. It can be observed that the sum rate achieved by PS remains nearly constant as the number of users increases. This behavior arises because the pinching beamformer is optimized independently in each time slot for the scheduled user, which allows constructive signal combining tailored to each user's location. In contrast, under PM, all users share the same pinching beamformer. As the number of users increases, this shared beamformer becomes increasingly suboptimal for individual users, which causes the achievable sum rate to decrease. Moreover, the in-waveguide propagation loss under PM becomes more pronounced. This effect occurs because the pinching antennas must be distributed more broadly along the waveguide in order to serve users at different locations, which increases the average propagation distance within the waveguide and exacerbates attenuation.

\section{Conclusion}
This work investigated the potential of incorporating PAs into tri-hybrid beamforming. Unlike conventional reconfigurable antennas, PAs mitigate large-scale path loss through pinching beamforming, enabling tri-hybrid beamforming to achieve significantly higher performance than conventional hybrid beamforming under the same number of RF chains. The analysis also revealed the existence of an optimal number of PAs, underscoring the importance of proper PA configuration in practical PASS-based tri-hybrid beamforming deployments. Extending the proposed framework to multiuser settings is a natural direction for future work, which is currently ongoing. 
\bibliographystyle{IEEEtran}
\bibliography{mybib}
\end{document}